\documentclass [12pt]{article}
\usepackage {a4wide}
\usepackage {graphicx}
\usepackage {longtable}

\begin{document}
\begin{center}
\textbf{Determination of Upper Limit of Stars Mass Based
on Model of Expansive Nondecelerative Universe} 
\end{center}

\bigskip
\begin{center}
Jozef \v{S}ima and Miroslav S\'{u}ken\'{\i}k
\end{center}

\bigskip

\begin{center}
Slovak Technical University, Radlinsk\'{e}ho 9, 812 37 Bratislava, Slovakia
\end{center}

\begin{center}
\underline {sima@chtf.stuba.sk} 
\end{center}

\bigskip

\textbf{Abstract}. Incorporation of the Vaidya metric in the model of 
Expansive Nondecelerative Universe allows to localize the energy density of 
gravitational field that, subsequently, enables to determine the upper limit 
of stars mass. The upper limit decreases with cosmologic time and at present 
is close to 30-fold of our Sun mass.

\bigskip

\subsection*{Introduction}

\bigskip

In the stars two key contrary oriented energy-mass forces can be identified, 
namely the stream of radiation directing from the center to outside of a 
star, and the gravity attracting the star mass to its centre.

The central gravitational pressure $p_{g} $ created by a star is 
proportional to the square of its mass $m_{s} $

\begin{equation}
\label{eq1}
p_{g} \sim  m_{s}^{2} 
\end{equation}

Providing that, in general, the chief process of mass transformation and 
energy formation in stars is a standard reaction of thermonuclear synthesis 
[1], their radiation output $L$ relates to their mass according to

\begin{equation}
\label{eq2}
L \cong k.m_{s}^{3} 
\end{equation}

\noindent
where the value

\begin{equation}
\label{eq3}
k \cong 5 \times 10^{ - 65}W.kg^{ - 3}
\end{equation}

\noindent
is attributed to the constant $k$. It follows from relations (\ref{eq1}) and (\ref{eq2}) 
that increasing the star mass, its radiation output grows more extensively 
than its gravitational pressure. A consequence is that star becomes unstable 
from the viewpoint of gravity at a certain limit mass $m_{s\left( {max} 
\right)} $. According to the literature data [1]

\begin{equation}
\label{eq4}
m_{max} \cong 20m_{Sun} 
\end{equation}

\noindent
where $m_{Sun} $ is the present mass of our Sun. It is, however, worth 
mentioning that stars of a mass as high as

\begin{equation}
\label{eq5}
m_{s} \cong 60m_{Sun} 
\end{equation}

\noindent
have been observed by astronomers. One of the reasons of a given discrepancy 
may lie in uncertainty of the constant $k$ value determination, further is 
an actual cosmological time of the star being observed (see relation (\ref{eq8})).

\bigskip

\subsection*{Results and discussion}

\bigskip

In the model of Expansive Nondecelerative Universe (ENU) the density of 
gravitational energy is in weak field conditions localizable, independent on 
the system of coordinates (it depends on the radial distance only), and can 
be expressed as [2]

\begin{equation}
\label{eq6}
\varepsilon _{g} = - \frac{{R.c^{4}}}{{8\pi .G}} = - \frac{{3m.c^{2}}}{{4\pi 
.a.r^{2}}}
\end{equation}

\noindent
where $\varepsilon _{g} $ is the density of gravitational energy emitted by 
a body with the mass \textit{m} at the distance \textit{r}, \textit{R} 
denotes the scalar curvature (contrary to a more frequently used 
Schwarzschild metric, in the Vaidya metric \textit{R $ \ne $} 0 also outside 
the body), and $a$ is the gauge factor. It is obvious that for a star to be 
stable, the absolute value of its gravitational output must be higher (equal 
in limiting case) than the value of its radiation output. It means

\begin{equation}
\label{eq7}
k.m_{max}^{3} \le \frac{{d}}{{dt}}\int {\frac{{R.c^{4}}}{{8\pi .G}}dV} 
\end{equation}

\noindent
and, in turn

\begin{equation}
\label{eq8}
k.m_{max}^{3} \le \frac{{m_{max} .c^{2}}}{{t_{c}} }
\end{equation}

\noindent
where $t_{c} $ is the cosmological time [3] reaching at present

\begin{equation}
\label{eq9}
t_{c} \cong 4.5 \times 10^{17}s
\end{equation}

Relations (\ref{eq3}), (\ref{eq8}) and (\ref{eq9}) lead to

\begin{equation}
\label{eq10}
m_{max} \le \sqrt {\frac{{c^{2}}}{{k.t_{c}} }} \cong 30m_{Sun} 
\end{equation}

The above result is in good accord with observations. Of course, observing 
(at the time being) very distant stars, i.e. seeing them as they were in a 
much shorter $t_{c} $, such stars may appear as more massive ones since the 
shorter $t_{c} $, the higher $m_{s\left( {max} \right)} $ (see relation 10).

We suppose there should be a dependence of the upper mass of stars and their 
cosmologic red shift.

\bigskip

\subsection*{References}

\bigskip

1. I.L. Rozental, \textit{Adv. Math. Phys. Astronomy, 31} (1986) 241 (in 
Czech)

\noindent
2. J. \v{S}ima, M. S\'{u}ken\'{\i}k, \textit{General Relativity and Quantum Cosmology,} 
gr-qc/9903090 

\noindent
3. V. Skalsk\'y, M. S\'{u}ken\'{\i}k, \textit{Astrophys. Space Sci., 190} (1992) 145

\end{document}